\documentclass[a4paper,12pt]{article}

\usepackage[dvips]{graphicx}

\providecommand{\delopenpsi}[1][]{\ensuremath{{\scriptstyle\Delta}\psi_{#1}}}

\providecommand{\Ngamma}{\ensuremath{N_{\gamma}}}
\providecommand{\Egamma}[1][]{\ensuremath{E_{\gamma\,#1}}}

\providecommand{\Minv}[1][]{\ensuremath{M_{inv\,#1}}}
\providecommand{\Mmis}[1][]{\ensuremath{M_{mis\,#1}}}

\providecommand{\chisq}[1][]{\ensuremath{\chi_{#1}^{2}}}

\providecommand{\etaeegg}{\ensuremath{{\phi\rightarrow{\eta}e^+e^-},
\ {\eta\rightarrow\gamma\gamma}}}

\providecommand{\phietag}{\ensuremath{{\phi\rightarrow\eta\gamma}}}

\providecommand{\etappg}{\ensuremath{{\eta\rightarrow\pi^+\pi^-\gamma}}}

\providecommand{\etappp}{\ensuremath{{\eta\rightarrow\pi^+\pi^-\pi^0}}}

\providecommand{\etapz}{\ensuremath{{\eta{\rightarrow}3\pi^0}}}

\providecommand{\phietaee}{\ensuremath{{\phi\rightarrow{\eta}e^+e^-}}}

\providecommand{\phietamm}{\ensuremath{{\phi\rightarrow{\eta}\mu^+\mu^-}}}

\providecommand{\etaeeg}{\ensuremath{{\eta{\rightarrow}e^+e^-\gamma}}}

\providecommand{\pioeeg}{\ensuremath{{\pi^0{\rightarrow}e^+e^-\gamma}}}

\providecommand{\phippp}{\ensuremath{{\phi\rightarrow\pi^+\pi^-\pi^0}}}

\providecommand{\etagg}{\ensuremath{{\eta\rightarrow\gamma\gamma}}}
\providecommand{\piogg}{\ensuremath{{\pi^0\rightarrow\gamma\gamma}}}

\providecommand{\MeV}{\mbox{MeV}}

\providecommand{\pb}{\mbox{pb}}
\providecommand{\etaeefourtr}{\ensuremath{\phi\to\eta{e^+e^-}, \eta\to\pi^+\pi^-\pi^0}}
\providecommand{\etaxee}{\ensuremath{\eta\to{X}e^+e^-}}
\providecommand{\etaxg}{\ensuremath{\eta\to{X}\gamma}}
\providecommand{\etappee}{\ensuremath{\eta\to\pi^+\pi^-e^+e^-}}
\providecommand{\etaeeee}{\ensuremath{\eta\to e^+e^-e^+e^-}}

\title
{\Large
\bf \boldmath
Study of conversion decays \phietaee, \etaeeg\  and \etappee\  at
\mbox{CMD-2} 
}

\author{
\parbox[c]{\textwidth}{\centering
R.R.~Akhmetshin\footnote{Budker Institute of Nuclear Physics,
Novosibirsk, 630090, Russia}, 
E.V.~Anashkin\footnotemark[1],
M.~Arpagaus\footnotemark[1],
V.M.~Aulchenko\footnotemark[1]\footnote{Novosibirsk State University, 
Novosibirsk, 630090, Russia}, 
V.Sh.~Banzarov\footnotemark[1],
L.M.~Barkov\footnotemark[1] \footnotemark[2],
N.S.~Bashtovoy\footnotemark[1], 
A.E.~Bondar\footnotemark[1] \footnotemark[2],
D.V.~Bondarev\footnotemark[1],
A.V.~Bragin\footnotemark[1], 
D.V.~Chernyak\footnotemark[1], 
S.I.~Eidelman\footnotemark[1] \footnotemark[2],
G.V.~Fedotovitch\footnotemark[1] \footnotemark[2], 
N.I.~Gabyshev\footnotemark[1], 
A.A.~Grebeniuk\footnotemark[1],
D.N.~Grigoriev\footnotemark[1], 
V.W.~Hughes\footnote{Yale University, New Haven, CT 06511, USA},
F.V.~Ignatov\footnotemark[1] \footnotemark[2],
P.M.~Ivanov\footnotemark[1], 
S.V.~Karpov\footnotemark[1],
V.F.~Kazanin\footnotemark[1] \footnotemark[2], 
B.I.~Khazin\footnotemark[1] \footnotemark[2],
I.A.~Koop\footnotemark[1],
P.P.~Krokovny\footnotemark[1] \footnotemark[2], 
L.M.~Kurdadze\footnotemark[1] \footnotemark[2], 
A.S.~Kuzmin\footnotemark[1] \footnotemark[2],
M.~Lechner\footnotemark[1],
I.B.~Logashenko\footnotemark[1], 
P.A.~Lukin\footnotemark[1],
K.Yu.~Mikhailov\footnotemark[1] \footnotemark[2],
I.N.~Nesterenko\footnotemark[1], 
V.S.~Okhapkin\footnotemark[1],
A.V.~Otboev\footnotemark[1],
E.A.~Perevedentsev\footnotemark[1] \footnotemark[2],
A.S.~Popov\footnotemark[1] \footnotemark[2],
T.A.~Purlatz\footnotemark[1] \footnotemark[2], 
N.I.~Root\footnotemark[1] \footnotemark[2],
A.A.~Ruban\footnotemark[1],
N.M.~Ryskulov\footnotemark[1],
A.G.~Shamov\footnotemark[1], 
Yu.M.~Shatunov\footnotemark[1],
B.A.~Shwartz\footnotemark[1] \footnotemark[2],
A.L.~Sibidanov\footnotemark[1] \footnotemark[2],
V.A.~Sidorov\footnotemark[1], 
A.N.~Skrinsky\footnotemark[1],
V.P.~Smakhtin\footnotemark[1],
I.G.~Snopkov\footnotemark[1], 
E.P.~Solodov\footnotemark[1] \footnotemark[2],
P.Yu.~Stepanov\footnotemark[1],
A.I.~Sukhanov\footnotemark[1], 
J.A.~Thompson\footnote{University of Pittsburgh, Pittsburgh, PA 15260, 
USA},
V.M.~Titov\footnotemark[1],
A.A.~Valishev\footnotemark[1], 
Yu.V.~Yudin\footnotemark[1],
S.G.~Zverev\footnotemark[1]
}
}

\date{}

\begin{document}

\maketitle

\begin{abstract}
Using 15.1 $\mbox{pb}^{-1}$ of data collected by \mbox{CMD-2}
in the 
$\phi$-meson energy range,
the branching ratios of the following conversion decays
have been measured:
$$
B(\phi\to\eta e^+e^-)      = (1.14\pm0.10\pm0.06)\cdot10^{-4},
$$
$$
B(\eta\to e^+e^-\gamma)    = (7.10\pm0.64\pm0.46)\cdot10^{-3},
$$
$$
B(\eta\to\pi^+\pi^-e^+e^-) = (3.7^{+2.5}_{-1.8}\pm0.3)\cdot10^{-4}.
$$
The upper limits for the following rare conversion decays
have been obtained at the 90\% confidence level:
$$
B(\phi\to\eta\mu^+\mu^-)   < 9.4\cdot10^{-6},
$$
$$
B(\eta\to e^+e^-e^+e^-)    < 6.9\cdot10^{-5}.
$$
\end{abstract}

\section{Introduction}

Conversion decays of low mass vector mesons 
$\rho,\omega,\phi \to \pi^0(\eta) + e^+e^-(\mu^+\mu^-)$ 
as well as Dalitz decays $\pi^0 (\eta) \to e^+e^-\gamma,
\eta \to \mu^+\mu^-\gamma$ have been extensively discussed 
\cite{faes} as one of the main background sources for experiments 
in which the yield of dileptons was measured in  heavy ion collisions
\cite{ceres_taps,helios}.  

By studying spectra of the invariant mass of a lepton pair
$\Minv(l^+l^-)$ in such decays one can determine a
so called transition form factor 
as a function of momentum transfer as well as the corresponding
branching ratio   and test predictions of 
various theoretical models, from standard vector meson dominance
(VMD) to calculations on the lattice
\cite{faes,landsberg,bramon,lattice}.


The experimental information on such decays is rather scarce \cite{pdg}.
While for the $\omega$ meson both possible conversion decays into 
$\pi^0$ were observed, $\omega \to \pi^0 \mu^+\mu^-$ \cite{lande} and
$\omega \to \pi^0 e^+e^-$ \cite{nd1},
only a few events of the decay $\phi \to \eta e^+e^-$ were previously
detected \cite{nd2}. The Dalitz decay of the $\pi^0$ was well 
studied before \cite{pdg}, but the situation with the Dalitz decay     
$\eta \to e^+e^-\gamma$ is much worse: only one experiment reported a 
measurement of its branching ratio \cite{jane}.

A large data sample of the $\phi$ mesons collected by \mbox{SND} and
\mbox{CMD-2} detectors at the \mbox{VEPP-2M} electron-positron
collider  dramatically changed the situation: both groups reliably 
measured the branching ratio of the decays
$\phi \to \eta e^+e^-$ and $\eta \to e^+e^-\gamma$ \cite{e1,e2}. 
\mbox{CMD-2} has recently published results of the first observation of 
the $\phi \to \pi^0 e^+e^-$ decay \cite{pioeegg_cmd2}.

This work is devoted to the determination of the branching ratio for the
conversion decays \phietaee, \etaeeg\  and \etappee\  using the
complete data sample available at \mbox{CMD-2}.
The upper limits for the branching ratios of the conversion decays 
\phietamm\  and \etaeeee\  were also obtained.

\section{Experiment}

The general purpose detector \mbox{CMD-2} installed at the
\mbox{VEPP-2M} $e^+e^-$ collider \cite{vepp} has been described in
detail elsewhere \cite{cmddec}.

It consists of a cylindrical drift chamber (DC) and double-layer
multiwire proportional Z-chamber, both also used for the trigger, and
both inside a thin (0.38 $X_0$) superconducting solenoid with a field
of 1T.
The momentum resolution of the DC is equal to
$\sigma_p / p = (\sqrt{90 \cdot (p(GeV))^2+7})~\%$ .
The accuracy in the measurement of polar and azimuthal angles is
$\sigma_{\theta}=1.5 \cdot 10^{-2}$ and $\sigma_{\phi} = 7 \cdot
10^{-3}$ radians respectively.

The barrel calorimeter with a thickness of 8.1$X_0$ is placed outside
the solenoid and consists of 892 CsI crystals.
The energy resolution for photons is about 9\% in the energy range
from 50 to 600 MeV.
The angular resolution is of the order of 0.02 radians.

The end-cap calorimeter placed inside the solenoid consists of 680 BGO
crystals.
The thickness of the calorimeter for normally incident particles is
equal to 13.4$X_0$.
The energy and angular resolution varies from 8\% to 4\% and from 0.03
to 0.02 radians respectively for the photon energy in the range
100 to 700 MeV.
Both barrel and end-cap calorimeters cover a solid angle of
0.92$\times4\pi$ steradians.

The experiment was performed in the $\phi$ meson energy range
(985-1060 MeV). The integrated luminosity collected
during the runs of 1993 (PHI93), 1996 (PHI96) and 1998 (PHI98)
was 1.5, 2.1 and 11.5~$\pb^{-1}$ respectively, so that our
analysis is based on the data sample corresponding to
15.1~$\pb^{-1}$.

\section{\boldmath Data analysis of decay \phietaee}

\subsection{\boldmath General approach}

Three different decay modes of the $\eta$ have been used for this analysis:
$\eta \to 2\gamma, 3\pi^0, \pi^+\pi^-\pi^0$. 
In the first and third cases event selection was performed using
kinematic reconstruction based on energy-momentum conservation. 
For each decay mode of the $\eta$ the number of detected events due to
the decay \phietaee\  is given by: 
\begin{equation}
\label{eq:n_phietaee}
N_{\phietaee} = N_{\phi} \cdot B(\phietaee) \cdot B(\eta\to{f}) \cdot
\varepsilon_{\phietaee},
\end{equation}
where $N_{\phi}$ is the total number of the produced $\phi$ mesons,
$B(\phietaee)$ is the branching ratio of the decay under study,
$B(\eta\to{f})$ is the product of the branching ratios for all 
intermediate decays involved in the decay chain leading to some
particular final state $f$ used for selection 
and $\varepsilon_{\phietaee}$ is the corresponding detection efficiency.

One of the most important backgrounds to the studied process comes
from events of the decay \phietag\  followed by the $\gamma$
conversion in the material in front of DC (hereafter referred to as 
``conversion'').
Since the DC resolution is not sufficient to separate events with the
conversion in the material from those due to conversion decays at the
interaction point, the contribution of this background was calculated
from the simulation: 
\begin{equation}
\label{eq:n_phietaee_phietag}
N_{\phietag} = N_{\phi} \cdot B(\phietag) \cdot B(\eta\to{f})
\cdot \varepsilon_{\phietag},
\end{equation}
where $\varepsilon_{\phietag}$ is the detection efficiency of the
\phietag\  decay with the $\gamma$ conversion in the material.

The total number of observed events is a sum of the two contributions
above: 
\begin{equation}
\label{eq:n_phietaee_exp}
N_{\phietaee}^{exp} = N_{\phietaee} + N_{\phietag}.
\end{equation}

The following expression for the branching ratio of the decay
\phietaee\  can be obtained from (\ref{eq:n_phietaee}),
(\ref{eq:n_phietaee_phietag}) and (\ref{eq:n_phietaee_exp}): 
\begin{equation}
\label{eq:b_phietaee}
\begin{array}{c}
B(\phietaee) = \frac{\displaystyle\mathstrut N_{\phietaee}^{exp}}
{\displaystyle\mathstrut N_{\phi}\cdot
B(\eta\to{f})\cdot\varepsilon_{\phietaee}} - \\
-
B(\phietag) \cdot \left( \frac{\displaystyle\mathstrut \varepsilon_{\phietag}} 
{\displaystyle\mathstrut \varepsilon_{\phietaee}} \right).
\end{array}
\end{equation}

Detection efficiencies were determined from Monte Carlo simulation
\cite{cmd2sim}. To take into account effects not properly reproduced 
by the simulation, the detection efficiencies were if necessary 
corrected using the information obtained from ``clean'' experimental
events. For example, for all conversion decays into an 
$e^+e^-$ pair the angle between
the electron and positron is small. Therefore, to determine the
reconstruction efficiency for events with a small opening angle between the 
tracks, a sample of experimental events of the process
$\phi \to \pi^+\pi^-\pi^0, \pi^0 \to e^+e^-\gamma$ was used.   

For the analysis of the decay mode $\eta \to 3\pi^0$ in which
$e/\pi$ separation described in detail elsewhere
\cite{pioeegg_cmd2,rho4pi} was applied, the corresponding correction
to the detection efficiency was obtained from the sample of 
experimental events of the processes $e^+e^- \to e^+e^-\gamma$ and 
$\phi \to \pi^+\pi^-\pi^0, \pi^0 \to 2\gamma$.
In this procedure the distributions of the parameter $E_{CsI}/p$ were 
studied for both particle types $(e/\pi)$ and signs $(+/-)$, where 
$E_{CsI}/p$ is the ratio of the energy deposition in the CsI calorimeter 
and the momentum of the particle with a track matching the cluster in CsI.
For each event the probability $W_{e^+e^-}^{e/\pi}$ for two tracks to be 
an $e^+e^-$ pair is determined from the probability density functions 
for the parameter $E_{CsI}/p$ as a function of particle momentum. 

In all cases the number of the $\phi$ mesons was determined 
from a process with a similar final state, so that some of 
systematic uncertainties cancel. It was specially checked that
the difference in the energy dependence of the process under study
and the normalization one produced negligible effect on the
final results. 

\subsection{\boldmath Selection of \phietaee\  by \etagg\  mode}
\label{sec:etaeegg}

In this process the final state contains two charged particles
and two photons,  
\mbox{$B(\eta\to{f})$}=\mbox{$B(\etagg)$}.

The following selection criteria were used for the
decay \phietaee:
\begin{itemize}
\item the angle between two tracks $\delopenpsi < 0.5$;
\item the number of photons taken for the kinematic 
reconstruction $\Ngamma^{KREC} = 2$, if the number of photons in an event
$N_{\gamma} > 2$, the combination with the best $\chisq$ was chosen;
\item the energy of each photon  $50<\Egamma<490~\MeV$.
\end{itemize}

\begin{figure}[ht]
\centering
\includegraphics[width=0.47\textwidth]{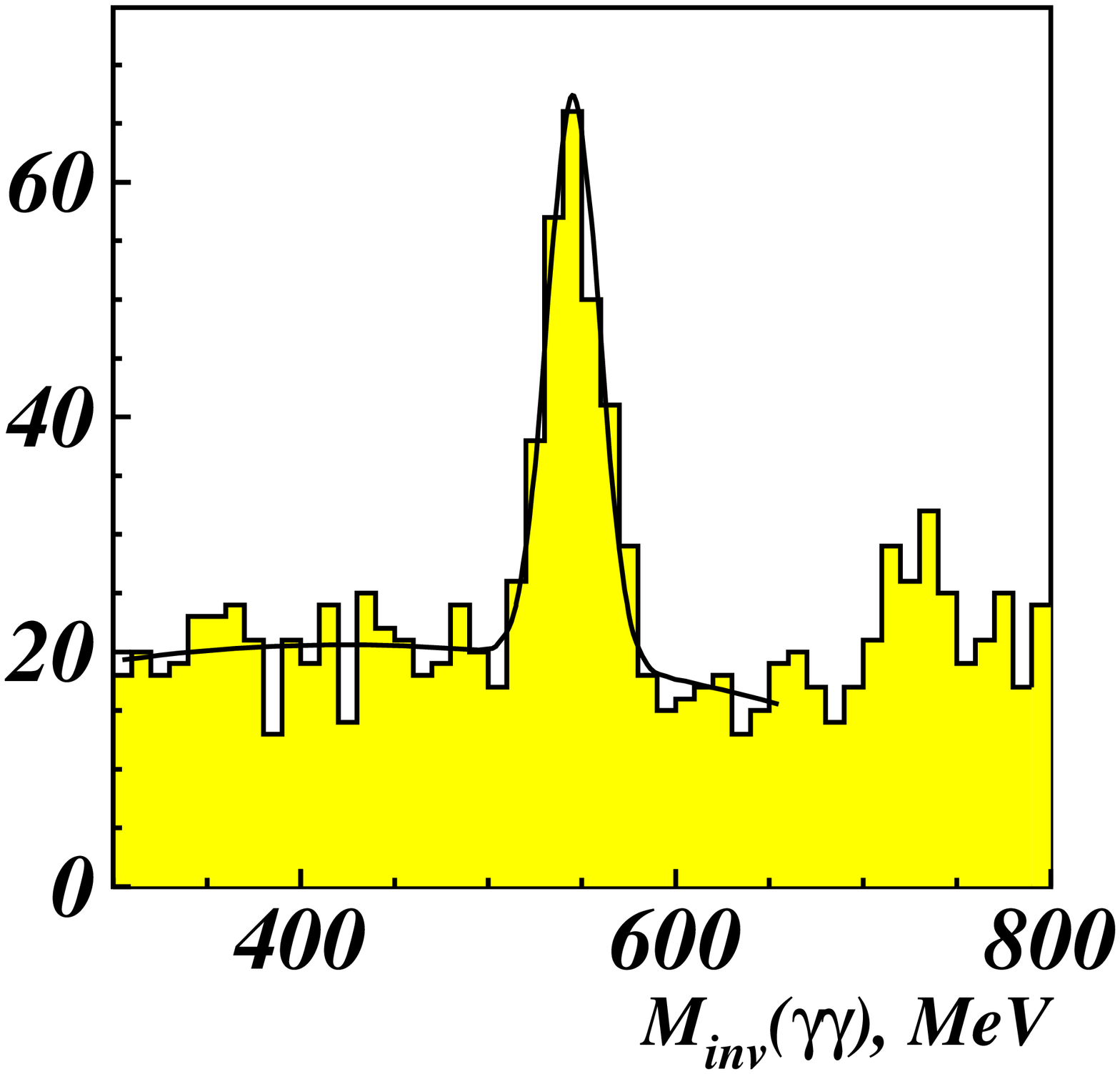}
\vspace{-0.5cm}
\caption{
Invariant mass $\Minv(\gamma\gamma)$ for \phietaee, \etagg}
\label{fig:etaeegg_phi98}
\end{figure}

Figure~\ref{fig:etaeegg_phi98} shows the distribution of the invariant
mass $\Minv(\gamma\gamma)$ for the selected events of the PHI98 run. 
A clear signal is observed at the $\eta$ meson mass which can be
fit with a Gaussian, its variance   taken from simulation.
The number of events in the peak is $167\pm18$ and the expected
number of ``conversion'' events is $23 \pm 2$.
The detection efficiencies from simulation are 
$\varepsilon_{\phietaee}=22.9\%$ and $\varepsilon_{\phietag}=3.1\cdot10^{-4}$.

The selection criteria for the the normalization process 
\phietag, \etappg\  were described elsewhere \cite{pioeegg_cmd2}. 
The number of detected events for it is
$1858\pm58$ and the detection efficiency is $20.8\%$.
The branching ratio of
the decay \phietaee\  is  $(1.10\pm0.15)\cdot10^{-4}$.

Experimental data of the PHI93 and PHI96 runs were processed
similarly.

The total detected number of the  \phietaee, \etagg\
candidates is $214\pm20$  with an expected ``conversion'' background 
of $31\pm2$ events.

The main sources of systematic uncertainties 
and theirs contributions to the error of the branching ratio
are listed below:
\begin{itemize}
\item A limited sample of simulated events used to determine detection 
efficiencies --- 2.6\%.
\item Statistical errors of the parameters in the correction for a small 
opening angle  --- 3.4\%.
\item Dependence on the transition form factor model --- 4.5\%.
\item The shape of the invariant mass distributions used in the fits
--- 2.4\%.
\item The branching ratios of intermediate decays, mainly
$B(\phietag)$ and $B(\etappg)$ --- 3.8\%.
\item Inaccurate knowledge of the material thickness for the calculations
of the ``conversion''  --- 1.3\%.
\end{itemize}
Two first sources of the error are of statistical nature and
since their values were obtained for each run separately, they
are run-to-run uncorrelated. Their total contribution in each
run was quadratically added to the statistical error and 
the values of the branching ratio in different runs were averaged.
  
The average value of the branching ratio of the \phietaee\  decay is
\mbox{$(1.13\pm0.14\pm0.07)\cdot10^{-4}$} where the first error is
statistical including the uncorrelated systematic errors described above
and the second one is a systematic correlated error common for all runs.

\subsection{\boldmath Selection of \phietaee\  by \etapz\  mode}

In this case the final state contains two charged particles
and 6 photons and
$B(\eta\to{f})~=~B(\etapz)\cdot(B(\piogg))^3$.

\begin{figure}[ht]
\includegraphics[width=0.47\textwidth]{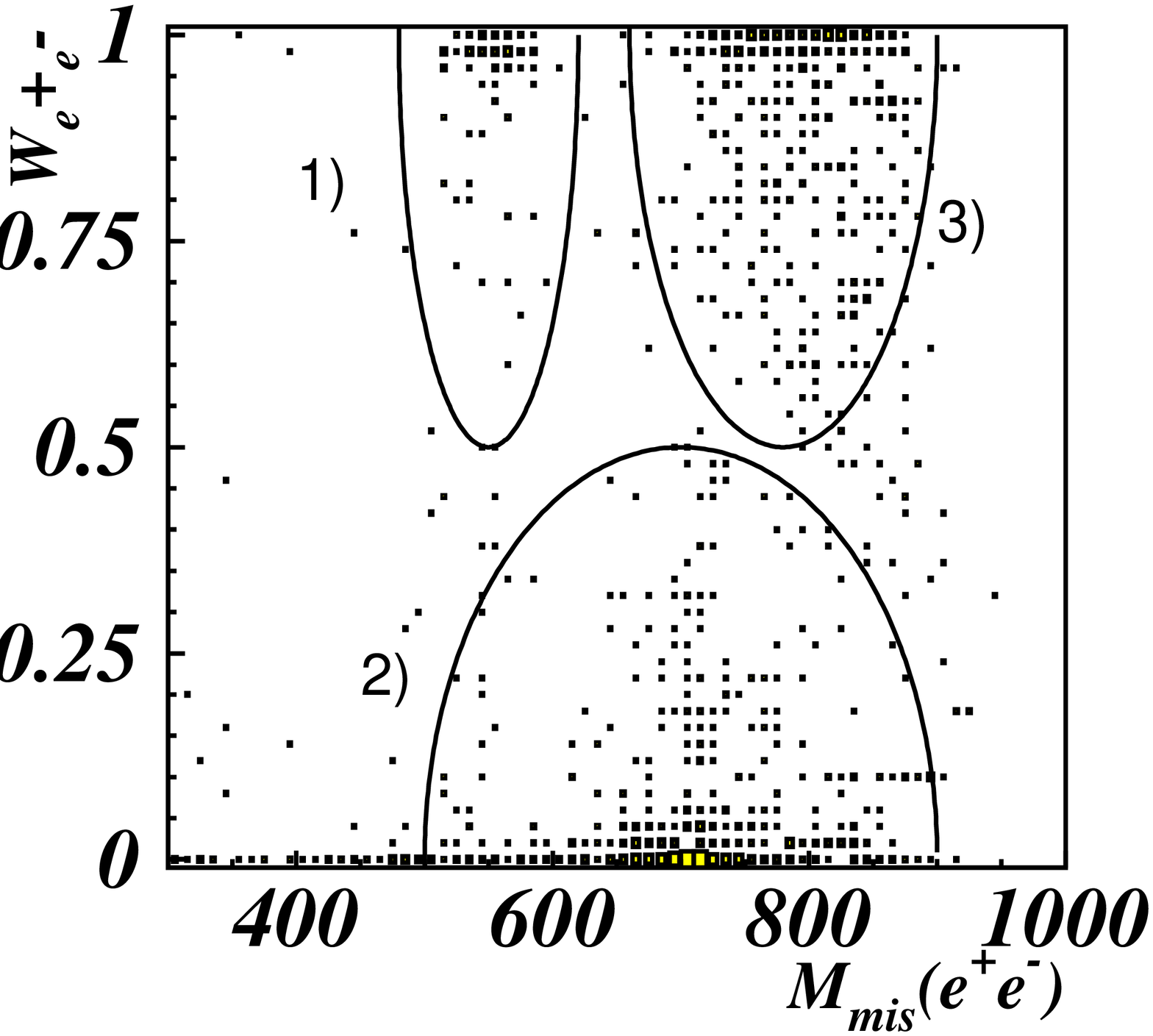}
\hfill
\includegraphics[width=0.47\textwidth]{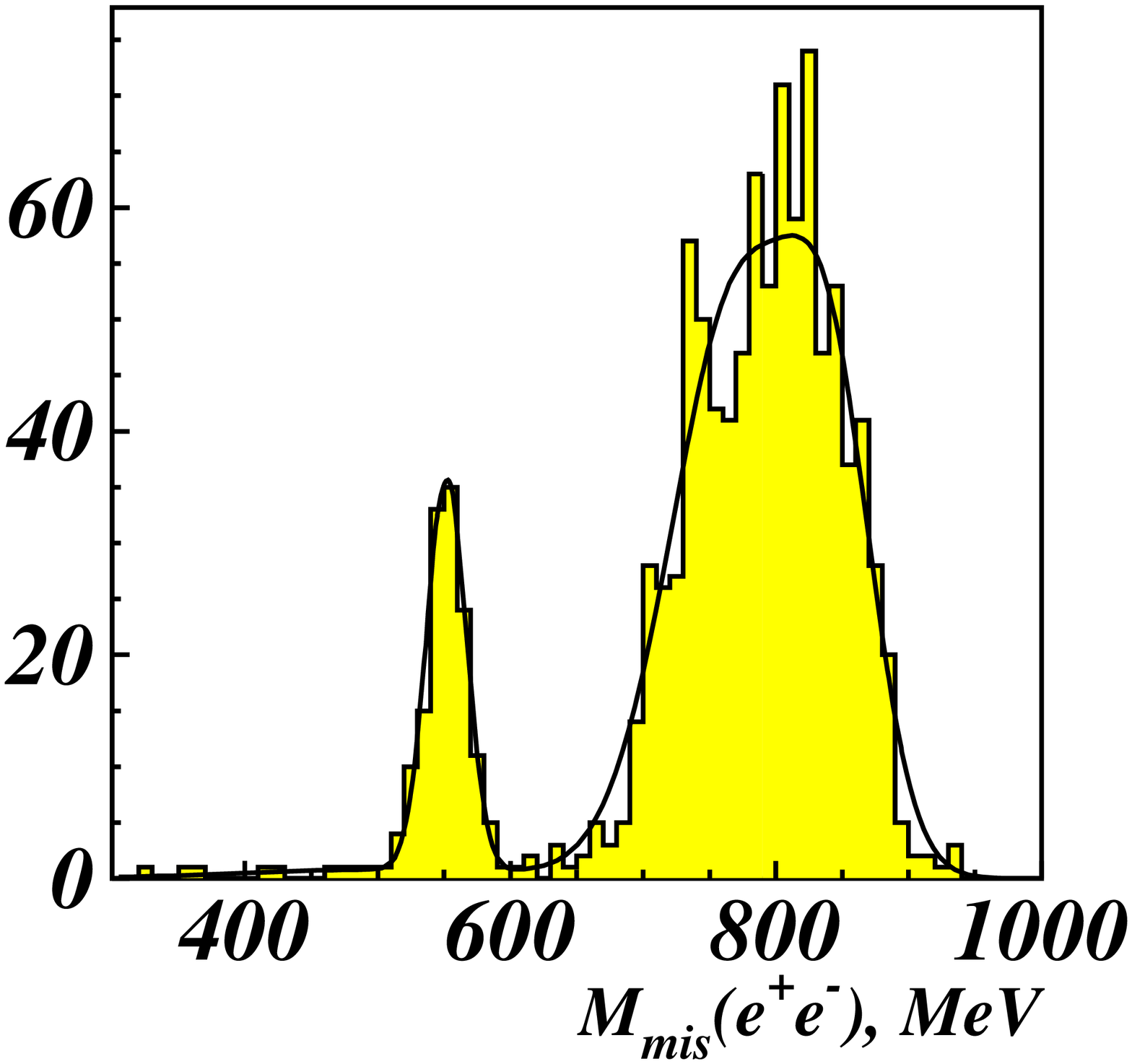}
\\
\parbox[t]{0.47\textwidth}{\vspace{-0.5cm}
\caption{Distribution of the probability $W_{e^+e^-}^{e/\pi}$ versus
the missing mass $\Mmis(e^+e^-)$ for \phietaee, \etapz}
\label{fig:ecltpt_vs_etaeeppp_phi98}
}
\hfill
\parbox[t]{0.47\textwidth}{\vspace{-0.5cm}
\caption{Missing mass $\Mmis(e^+e^-)$ for \phietaee, \etapz}
\label{fig:etaeeppp_phi98}
}
\end{figure}

The following selection criteria were used for the decay \phietaee:
\begin{itemize}
\item $\delopenpsi < 0.5$;
\item $\Ngamma \ge 4$;
\item $\Egamma^{max} < 490~\MeV$.
\end{itemize}
The event by event $e/\pi$-separation described previously was employed,
so that $W_{e^+e^-}^{e/\pi}$ was determined for each event.
Figure~\ref{fig:ecltpt_vs_etaeeppp_phi98} shows the distribution of the
probability $W_{e^+e^-}^{e/\pi}$ versus the missing mass of tracks 
$\Mmis(e^+e^-)$ for the PHI98 data after these cuts.
Events with an $e^+e^-$-pair have $W_{e^+e^-}^{e/\pi}\sim1$, while
for those with pions $W_{e^+e^-}^{e/\pi}\sim0$.
Events of the process \phietaee, \etapz\
(region 1 in Fig.~\ref{fig:ecltpt_vs_etaeeppp_phi98}) are separated from
events of the process $\phi\to K_S K_L$, $K_S\to\pi^+\pi^-\gamma$
(region 2 in Fig.~\ref{fig:ecltpt_vs_etaeeppp_phi98}) and from events of
the process $\phi\to K_S K_L$, $K_S\to\pi^0\pi^0$, $\pi^0\to e^+e^-\gamma$
(region 3 in Fig.~\ref{fig:ecltpt_vs_etaeeppp_phi98}).
It is clear that the following additional cut can efficiently separate
events with pions from those with electrons: 
\begin{itemize}
\item \mbox{$W_{e^+e^-}^{e/\pi}>0.5$}.
\end{itemize}


Figure~\ref{fig:etaeeppp_phi98} shows the distribution of the missing mass
$\Mmis(e^+e^-)$ for the PHI98 data after all cuts.
The left peak contains the events of the decay under study while the 
right peak (at $\sim800$~MeV) corresponds to $\phi\to K_S K_L$,
$K_S\to\pi^0\pi^0$, $\pi^0\to e^+e^-\gamma$ events.
The corrected detection efficiencies are 
$\varepsilon_{\phietaee}=19.3\%$ and
$\varepsilon_{\phietag}=3.9\cdot10^{-4}$.

From the fit the number of events in the left peak is $131\pm12$,
while the expected number of ``conversion'' events is
$23 \pm 12$. Normalizing to the same process $\phi \to \eta \gamma,
\eta \to \pi^+\pi^-\gamma$ as in subsection \ref{sec:etaeegg}, 
one obtains the
branching ratio of the decay \phietaee\  $(1.23\pm0.16)\cdot10^{-4}$.

The total  number of candidates for the process \phietaee, \etapz\  is
$158\pm13$ events with an expected ``conversion'' background of 
$28\pm2$ events.

In addition to the sources of systematic uncertainties described in the
previous subsection, there is a systematic uncertainty arising
from $e/\pi$ separation. 
Since this procedure is based on the
independent data sample and is therefore run-to-run uncorrelated, the
corresponding systematic uncertainty equal to 1.8\%
can be added to the statistical error of the branching ratio in each run.


The average value of the branching ratio of the \phietaee\  decay is
\mbox{$(1.21\pm0.14\pm0.09)\cdot10^{-4}$}.

\subsection{\boldmath Selection of $\phi\to\eta{e^+e^-}$  by
$\eta\to\pi^+\pi^-\pi^0$ mode} 
\label{sec:etaeefourtr}

For this process the final state contains four charged particles
and two photons,
$B(\eta\to{f})$~=~$B(\etappp)\cdot B(\piogg)$.

Two tracks with opposite charges and a smaller opening angle were 
assumed to be an $e^+e^-$ pair, while two others -- correspondingly
a pion pair.



\begin{figure}[ht]
\includegraphics[width=0.47\textwidth]{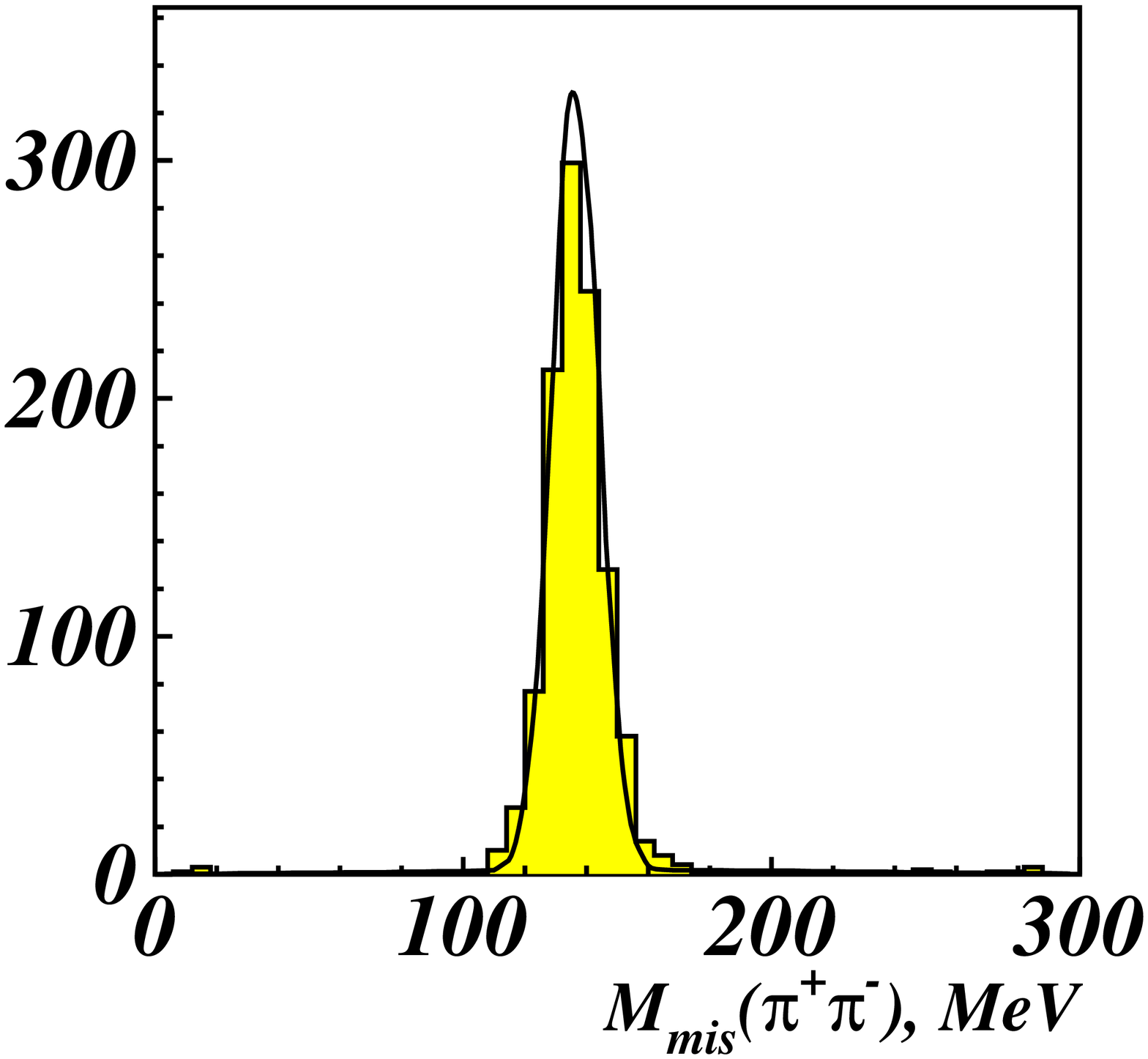}
\hfill
\includegraphics[width=0.47\textwidth]{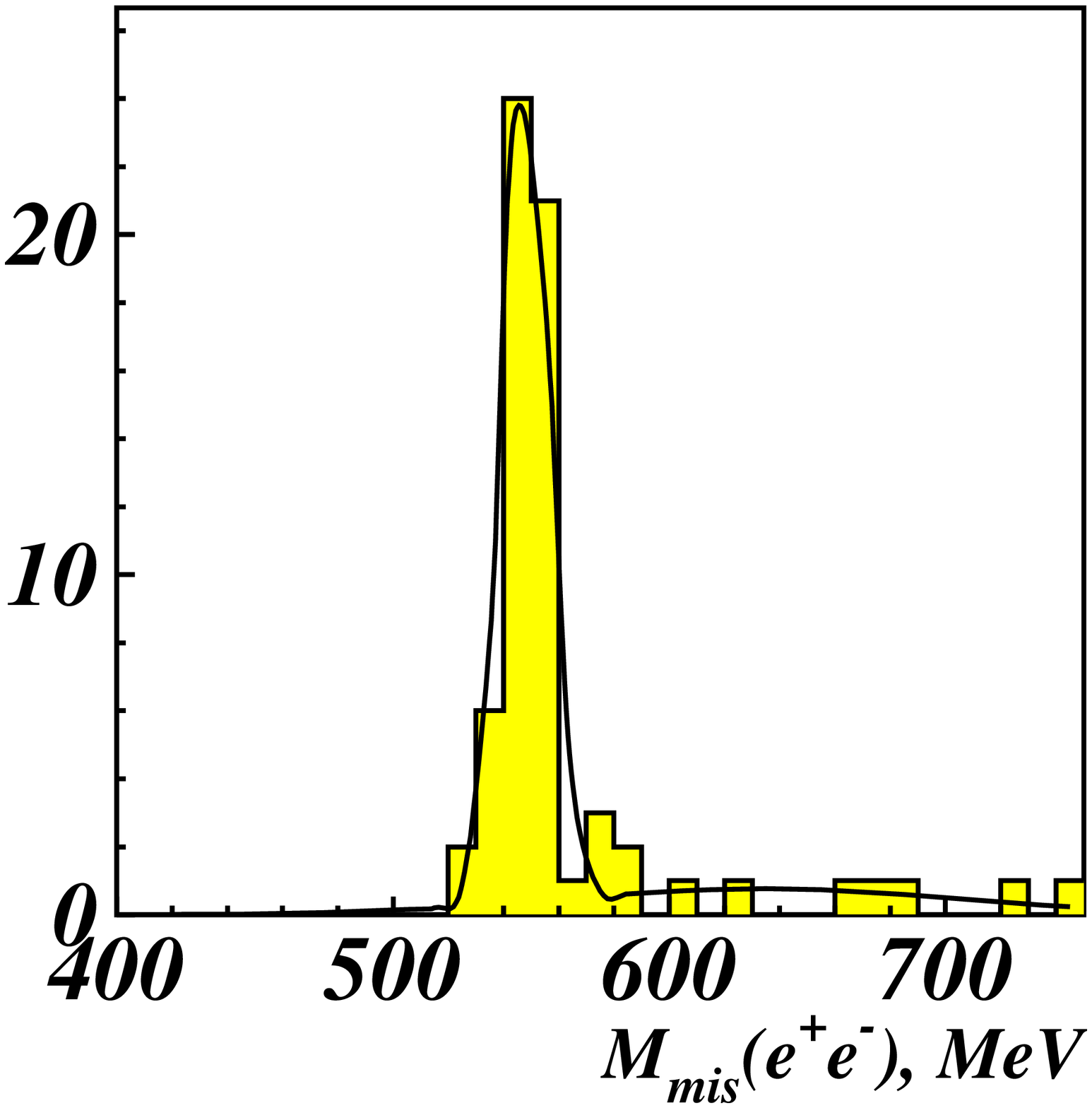}
\\
\parbox[t]{0.47\textwidth}{\vspace{-0.5cm}
\caption{Invariant mass $\Minv(e^+e^-\gamma)$ for the normalization
 process \phippp, \pioeeg} 
\label{fig:pipipieeg}
}
\hfill
\parbox[t]{0.47\textwidth}{\vspace{-0.5cm}
\caption{Missing mass $\Mmis(e^+e^-)$ for \phietaee, \etappp, \piogg}
\label{fig:etaeefourtr}
}
\end{figure}

The selection criteria for the normalization process 
\phippp, \pioeeg\  are:
\begin{itemize}
\item $\delopenpsi(e^+e^-) < 0.3$;
\item $0.5<\delopenpsi(\pi^+\pi^-) < 2.5$, to suppress $\phi{\to}K_SK_L$;
\item the number of photons $\Ngamma^{KREC} = 1$;
\item the photon energy $50<\Egamma < 250~\MeV$.
\end{itemize}
Figure~\ref{fig:pipipieeg} shows the distribution of the invariant mass
$\Minv(e^+e^-\gamma)$ for all the data after the cuts.
The detection efficiencies from simulation are 
$\varepsilon_{\pioeeg}=4.8\%$ and
$\varepsilon_{\piogg}=9.4\cdot10^{-5}$.
The number of events in the peak for this process is $1745\pm43$.

The selection criteria for the process under study 
\phietaee, \etappp, \piogg\
are listed below:
\begin{itemize}
\item $\delopenpsi(e^+e^-) < 0.3$;
\item $0.5<\delopenpsi(\pi^+\pi^-) < 2.5$;
\item the number of photons $\Ngamma^{KREC} = 2$;
\item the photon energy $50<\Egamma < 250~\MeV$;
\item $\Minv(e^+e^-\gamma) > 300~\MeV$, to suppress the Dalitz decay
of the $\pi^0$;
\item $\Minv(\gamma\gamma)< 250~\MeV$.
\end{itemize}
Figure~\ref{fig:etaeefourtr} shows the distribution of the missing mass
$\Mmis(e^+e^-)$ for all the data after the cuts.
The detection efficiencies are $\varepsilon_{\phietaee}=10.4\%$ and
$\varepsilon_{\phietag}=2.3\cdot10^{-4}$.
The number of \phietaee\  events in the peak   is $53\pm8$ and 
the expected number of ``conversion'' events is $11 \pm 1$.
The corresponding
branching ratio of the decay \phietaee\  is  
$(1.04\pm0.20\pm0.08)\cdot10^{-4}$.


Table \ref{tab:phietaee} summarizes results of the data processing for the 
\phietaee\  decay and shows the number of selected events, the expected 
number of ``conversion'' events as well as the
obtained branching ratios. 
It is clear that the values of $B(\phietaee)$ determined from various
decay modes of the $\eta$ are consistent within the errors and can be
averaged.  
The averaging procedure took into account that some of the sources of
the systematic error like e.g. the branching ratios of the
intermediate decays are common for three measurements.    

\begin{table}[ht]
\caption{Branching ratio of \phietaee\  decay}
\label{tab:phietaee}
\begin{center}
\begin{tabular}{lccc}
\hline
\hline
Mode & $N_{\phietaee}^{exp}$ &  $N_{\phietag}^{conv}$ &
$B(\phietaee)$, $10^{-4}$ \\
\hline
\etagg  & $214\pm20$ & $31\pm2$ & $1.13\pm0.14\pm0.07$ \\
\etapz  & $158\pm13$ & $28\pm2$ & $1.21\pm0.14\pm0.09$ \\
\etappp & $~53\pm~8$ & $11\pm1$ & $1.04\pm0.20\pm0.08$ \\
\hline
Total   & $425\pm25$ & $70\pm3$ & $1.14\pm0.10\pm0.06$ \\
\hline
\hline
\end{tabular}
\end{center}
\end{table}

\section{\boldmath Search for decay  $\phi \to \eta \mu^+\mu^-$}

Events with four tracks and two or more photons were analyzed to
search for the decay $\phi \to \eta\mu^+\mu^-$, $\eta \to
\pi^+\pi^-\pi^0$. For conversion decays into a $\mu^+\mu^-$ pair the
angle between muons is not necessarily small.

The analysis employed a kinematic fit taking into account 
energy-momentum conservation.
A pair of tracks with opposite charges and the missing mass closest
to the $\eta$ mass was assumed to be a muon pair.

The invariant mass of the photon pair was required to be near the
$\pi^0$ mass: $|M_{\gamma\gamma}-m_{\pi^0}| < 30$ MeV.

We restricted the ionization losses of the tracks (in arbitrary units):
$dE/dx < 2000 ( 1 + 2 | \cos \theta| ) \left( 1 +
  \frac{(85+15Q)^2}{\vec{p}^2}\right)$, where $Q$ and $\vec{p}$ are
the track charge and momentum (in $\mbox{MeV}/c$). This requirement
suppressed the background from the decay $\phi \to K^+K^-$ in which
products of kaon nuclear interactions scatter back to the drift
chamber and induce two extra tracks or one of the kaons decays via
the $K^{\pm}\to\pi^{\pm}\pi^+\pi^-$ channel, accompanied by fake
photons.

Then a pair of oppositely charged particles with the minimum
space angle $\psi_{min}$ between the tracks was looked for. Assuming
it  to be an $e^+e^-$ pair and taking a photon with the smaller
energy, the invariant mass $M_{e^+e^-\gamma_{2}}$ was
calculated. The requirements $\psi_{min} >$ 0.3 and
$M_{e^+e^-\gamma_{2}} >$ 170 MeV reduced the background from the
reaction $e^+e^- \to \omega\pi^0$ with the conversion decay of one of
the neutral pions.

To reject the background from the decay $\phi \to K_S
K_L$, $K_S \to \pi^+\pi^-$ and $K_L \to \pi^+\pi^-\pi^0$ we
excluded events in which at least one pair of tracks satisfies the 
conditions: $|M_{\pi^+\pi^-}-m_{K_S}| < 30 \,\mbox{MeV}/c^2$ and
$|P_{\pi^+\pi^-}-P_{K_S}| < 30 \,\mbox{MeV}/c$.
Here $M_{\pi^+\pi^-}$ and $P_{\pi^+\pi^-}$ are the
invariant mass and momentum of the pair calculated under the
assumption that particles in this pair are pions.

For the twelve events surviving the above conditions, the
distribution of the maximum impact parameter of tracks 
$d$ was analyzed.

The decay under study contributes to the region of small impact
parameters, $d < 0.3$ cm, while the main background reaction $\phi
\to K_S K_L$ has a broad distribution of $d$. We have detected
$\tilde{N}_{\eta\mu\mu} = 2$ candidate events with $d < 0.3$
cm. Using the number of $K_S K_L$ events $N_{K_S K_L} = 10$
observed in the region $d>0.3$ cm and the ratio
$N_{K_S K_L}(d<0.3)/N_{K_S K_L}(d>0.3) = 0.20 \pm 0.03$ 
obtained from simulation, the number of
background events was estimated to be $N_{bg} = 2.0 \pm 0.6$. 
This leads to an upper limit for  the number of signal events: 
$N_{\eta\mu\mu} < 3.9$ at 90\% CL. 
From simulation the detection efficiency was 
$\varepsilon = 9.1\%$. 
The upper limit for the decay
probability is $B(\phi \to \eta \mu^+\mu^-) < 9.4 \cdot 10^{-6}$ at
90\% CL.

\section{\boldmath Data analysis for $\eta$ conversion decays}

\subsection{\boldmath General approach}

Conversion decays of the $\eta$-meson were studied 
using the radiative decay of the $\phi$:
$e^+e^-\to\phi\to\eta\gamma$.

Similarly to the $\phi \to \eta e^+e^-$ decay, the number of detected events 
for the  conversion decay $\eta\to{X}e^+e^-$
is given by the following expression: 
\begin{equation}
\label{eq:n_etaxee}
N_{\etaxee} = N_{\phi} \cdot B(\phietag) \cdot B(\etaxee) \cdot
\varepsilon_{\etaxee},
\end{equation}
where 
$B(\etaxee)$ is the branching ratio of the $\eta$ decay under study
($X~=~\gamma, \pi^+\pi^-$) and $\varepsilon_{\etaxee}$ is the 
corresponding detection efficiency.

The contribution of the background caused by the conversion
in the material  was calculated from simulation: 
\begin{equation}
\label{eq:n_etaxee_etaxg}
N_{\etaxg} = N_{\phi} \cdot B(\phietag) \cdot B(\etaxg) \cdot
\varepsilon_{\etaxg},
\end{equation}
where $\varepsilon_{\etaxg}$ is the detection efficiency of the
$\eta$ decay with the $\gamma$ conversion in the material.

The total number of observed events is a sum of the two contributions
above: 
\begin{equation}
\label{eq:n_etaxee_exp}
N_{\etaxee}^{exp} = N_{\etaxee} + N_{\etaxg}.
\end{equation}

The following expression for the branching ratio of the decay
\phietaee\  can be obtained from (\ref{eq:n_etaxee}),
(\ref{eq:n_etaxee_etaxg}) and (\ref{eq:n_etaxee_exp}): 
\begin{equation}
\label{eq:b_etaxee}
\begin{array}{c}
B(\etaxee) = \frac{\displaystyle\mathstrut N_{\etaxee}^{exp}}
{\displaystyle\mathstrut N_{\phi}\cdot
B(\phietag)\cdot 
\varepsilon_{\etaxee}} - \\
-
B(\etaxg) \cdot \left( \frac{\displaystyle\mathstrut \varepsilon_{\etaxg}} 
{\displaystyle\mathstrut \varepsilon_{\etaxee}} \right).
\end{array}
\end{equation}

\subsection{\boldmath Selection of \etaeeg\  decay}

The final state contains two charged particles and two photons. 

\begin{figure}[ht]
\includegraphics[width=0.47\textwidth]{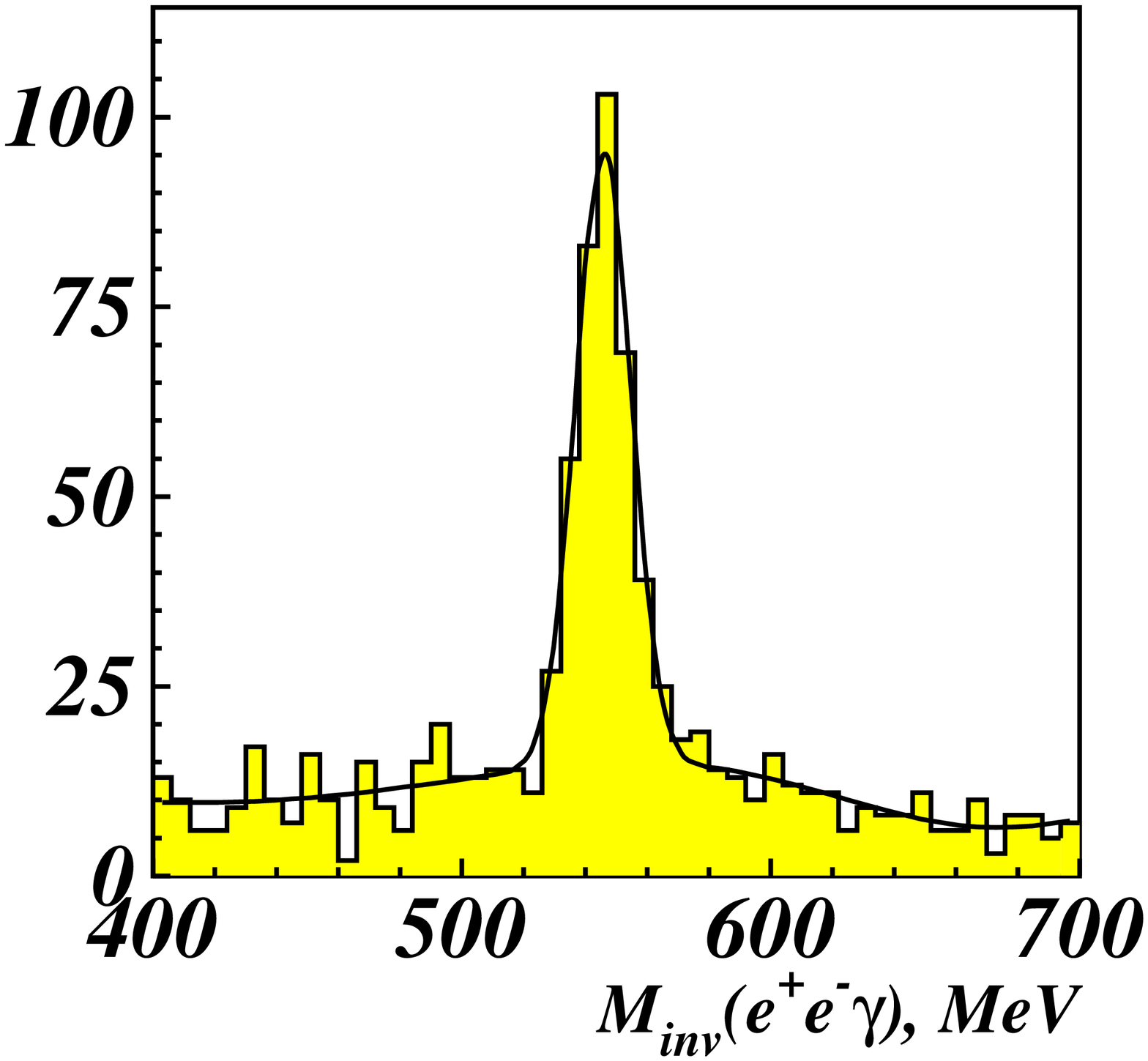}
\hfill
\includegraphics[width=0.47\textwidth]{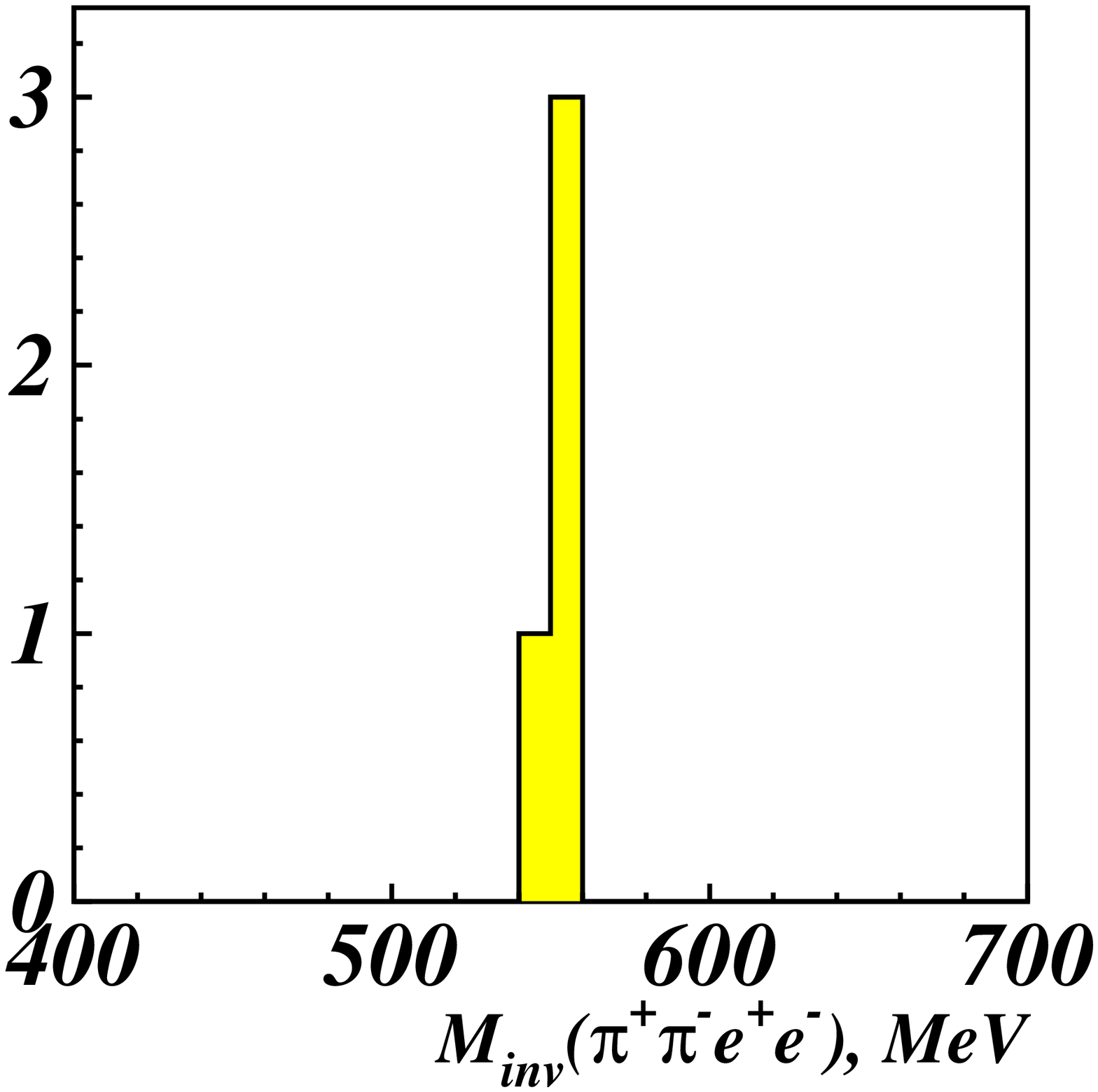}
\\
\parbox[t]{0.47\textwidth}{\vspace{-0.5cm}
\caption{Invariant mass $\Minv(e^+e^-\gamma)$ for \phietag, \etaeeg}
\label{fig:etageeg_phi98}
}
\hfill
\parbox[t]{0.47\textwidth}{\vspace{-0.5cm}
\caption{Invariant mass $\Minv(\pi^+\pi^-e^+e^-)$ for \phietag, \etappee}
\label{fig:etagppee}
}
\end{figure}

The normalization process used for the
decay \etaeeg\  was the same as for process \etaeegg\  (see 
subsection \ref{sec:etaeegg}). The selection criteria are also similar
to those for \etaeegg\  and include two additional cuts:
\begin{itemize}
\item $M_{inv}(\gamma\gamma) > 250$ MeV to suppress events of the
decay $\phi \to \pi^+\pi^-\pi^0$;
\item $|M_{inv}(\gamma\gamma) -547.5| >$ 30 MeV to suppress events
of the conversion decay $\phi \to \eta e^+e^-, \eta \to \gamma\gamma$.
\end{itemize}    
Figure~\ref{fig:etageeg_phi98} shows the distribution of the invariant
mass $\Minv(e^+e^-\gamma)$ for the selected events of the PHI98 run. 

A clear signal is observed at the $\eta$ meson mass which can be
fit with a Gaussian, its variance  taken from simulation.
The number of events in the peak is $303\pm21$ and the expected number of
``conversion'' events is $38\pm3$. The
detection efficiencies are $\varepsilon_{\etaeeg}=19.7\%$ and
$\varepsilon_{\etagg}=5.1\cdot10^{-4}$.

The total detected number of the  \phietag, \etaeeg\
candidates from all experimental runs is $374\pm24$  with an expected 
background of $51\pm3$ events.
The systematic errors are the same as for the process \etaeegg. 
The average value of the branching ratio of the \etaeeg\  decay is
\mbox{$(7.10\pm0.64\pm0.46)\cdot10^{-3}$}.

\subsection{\boldmath Selection of $\eta\to\pi^+\pi^-e^+e^-$  decay}

The final state contains four charged particles and one photon. 

The normalization process used for the decay \etappee\  was the same as 
for the process \etaeefourtr\  (see Section \ref{sec:etaeefourtr}).
The selection criteria were:
\begin{itemize}
\item $\delopenpsi(e^+e^-) < 0.5$;
\item the number of photons $\Ngamma^{KREC} = 1$;
\item $\Minv(e^+e^-\gamma) > 200~\MeV$ to suppress the Dalitz decay
of the $\pi^0$;
\item $\Mmis(e^+e^-)>600~\MeV$ to suppress events of the decay \phietaee.
\end{itemize}
Figure~\ref{fig:etagppee} shows the distribution of the invariant mass
$\Minv(\pi^+\pi^-e^+e^-)$ for all the data after applying these cuts.
The detection efficiencies are $\varepsilon_{\etappee}=5.04\%$ and
$\varepsilon_{\etappg}=4.73\cdot10^{-5}$.
Four events were detected  with the expected ``conversion''
background of 0.4. The background from events of the process 
\phietag, \etappp, \pioeeg\  with one lost
photon was estimated from simulation to be less than 0.1 event.
The systematic errors are similar to these for the
process \etaeefourtr. 
The branching ratio for the  decay \etappee\  is
$(3.7^{+2.5}_{-1.8}\pm0.3)\cdot10^{-4}$. 

\section{\boldmath Search for decay $\eta \to e^+e^-e^+e^-$}

Selection criteria similar to those for the decay $\eta \to e^+e^-\pi^+\pi^-$ 
were used for a search for the  decay \etaeeee:
\begin{itemize}
\item $\delopenpsi(e^+e^-)_{1,2} < 0.5$;
\item the number of photons $\Ngamma^{KREC} = 1$;
\item $\Minv(e^+e^-\gamma) > 200~\MeV$ to suppress the Dalitz decay
of the $\pi^0$.
\end{itemize}
After applying all these cuts no events survive with the  invariant mass
$\Minv(e^+e^-e^+e^-)$ near the $\eta$ mass. The
detection efficiencies are $\varepsilon_{\etaeeee}=17.5\%$ and
$\varepsilon_{\etaeeg}=9.4\cdot10^{-5}$. The normalization process
is the same as in the previous case.
The corresponding upper limit for the branching ratio of the decay 
\etaeeee\  is $6.9\cdot10^{-5}$ at the 90\% confidence level. 

\section{Discussion}

 
Table~\ref{tab:phiall} compares CMD-2 results for conversion decays
to those of other existing measurements as well as to the theoretical
predictions \cite{faes,eidelman,hashi}. Some scatter of the predictions
is caused by different assumptions about the transition form factors as
well as the values of the coupling constants used by the authors.   
The CMD-2 result for $B(\phietaee)$ supersedes the previous value
based on a small part of the total data sample collected by CMD-2
\cite{solodov,lechner} as well as the preliminary result of \cite{e2}. 
It can be seen that all results of this work are consistent with other 
measurements and are more precise. Their accuracy is comparable to that 
of the theoretical predictions.

\begin{table}[ht]
\begin{minipage}{\textwidth}
\caption{Branching ratios of conversion decays}
\label{tab:phiall}
\begin{center}
\begin{tabular}{lccc}
\hline
\hline
\hspace{-2mm} \begin{tabular}{c} Branching \\ ratio  \\ \end{tabular} & 
\begin{tabular}{c} \hspace{-5mm} $B(\phietaee)$, \hspace{-5mm} \\ $10^{-4}$ \\ \end{tabular} & 
\begin{tabular}{c} \hspace{-5mm} $B(\etaeeg)$,   \hspace{-5mm} \\ $10^{-3}$ \\ \end{tabular}  &
\begin{tabular}{c} \hspace{-5mm} $B(\etappee)$,  \hspace{-2mm} \\ $10^{-4}$ \\ \end{tabular}  \\
\hline
\hline
\hspace{-2mm} \begin{tabular}{l} Theory \\ \end{tabular} & 
$1.0-1.2$ & $6.5-6.8$ & $3.6$ \\
\hline
\hspace{-2mm} \begin{tabular}{l} CMD-2 \\ $[$this work$]$ \\ \end{tabular}  & 
$1.14\pm0.10\pm0.06$ & $7.10\pm0.64\pm0.46$ &
$3.7^{+2.5}_{-1.8}\pm0.3$ \\
\hspace{-2mm} \begin{tabular}{l} 
SND \cite{prsnd}\footnote{After SND results were published in \cite{prsnd}, its
authors came to the conclusion that the systematic uncertainties
for $B(\phietaee)$ and $B(\etaeeg)$ were underestimated by a factor of 
1.5-2 \cite{dimova}.}
 \\ \end{tabular} & 
$1.19\pm0.19\pm0.07$ & $5.15\pm0.62\pm0.39$ & $-$ \\
\hspace{-2mm} \begin{tabular}{l} Others \\ \end{tabular} & 
$1.3^{+0.8}_{-0.6}$ \cite{nd2} & $4.9\pm1.1$ \cite{jane} &
$13^{+12}_{~-8}$ \cite{gross} \\ 
\hline
\hline 
\end{tabular}
\end{center}
\end{minipage}
\end{table}

A search for the decay $\phi \to \eta\mu^+\mu^-$ has been performed
for the first time.  
The obtained upper limit 
$B(\phi \to \eta\mu^+\mu^-) < 9.4 \cdot 10^{-6}$ at 90\% CL
is a factor of 1.4-1.8 higher than the theoretical prediction of
$(5.3 -6.8) \cdot 10^{-6}$ \cite{faes,eidelman,hashi}.

Finally, the upper limit for the branching ratio of the decay 
\etaeeee\  has been obtained for the first time:
$B(\etaeeee) < 6.9 \cdot 10^{-5}$ at 90\% CL.
It is slightly above the theoretical estimate of $6.5\cdot10^{-5}$
based on the result of \cite{kroll}.

The applied cut on the angle between the tracks $\Delta \Psi < 0.5$
selects events with a rather small q$^2$ so that it is practically
impossible to study the momentum transfer dependence of the cross
section and therefore transition form factors. Such a study will
require a much larger data sample.
 
\section{Conclusions}

Using the total data sample of 15.1 pb$^{-1}$ collected by CMD-2 
in the c.m.energy range 985-1060 MeV, the following results
were obtained for various conversion decays of the $\phi$ and
$\eta$ mesons:
$$
B(\phietaee)~=~(1.14\pm0.10\pm0.06) \cdot 10^{-4},
$$
$$
B(\etaeeg)~=~(7.10\pm0.64\pm0.46) \cdot 10^{-3},
$$
$$
B(\etappee)~=~(3.7^{+2.5}_{-1.8}\pm0.3) \cdot 10^{-4},
$$
$$
B(\phietamm) < 9.4 \cdot 10^{-6} \mbox{~at~} 90\% \mbox{~CL},
$$
$$
B(\etaeeee) < 6.9 \cdot 10^{-5} \mbox{~at~} 90\% \mbox{~CL}.
$$
   
The authors are grateful to the staff of \mbox{VEPP-2M} for the
excellent performance of the collider, to all engineers and 
technicians who participated in the design, commissioning and operation
of \mbox{CMD-2}. 
We acknowledge stimulating discussions with R.A.~Eichler
and M.S.~Zolotorev.


\begin{thebibliography}{199}

\bibitem{faes}
A.~Faessler, C.~Fuchs and M.I.~Krivoruchenko,
Phys. Rev. {\bf C61} (2000) 035206.



\bibitem{ceres_taps}
G.~Agakichiev {\it et al.,}
Phys. Rev. Lett. {\bf 75} (1995) 1272.

\bibitem{helios}
M.~Masera, Nucl. Phys. {\bf A590} (1995) 93c.

\bibitem{landsberg}
L.~Landsberg, Phys. Rep. {\bf 128} (1985) 301.

\bibitem{bramon}
A.~Bramon, M.~Greco,
The Second DA$\Phi$NE Physics Handbook. INFN-Laboratori
Nazionali di Frascati. Edited by L.Maiani, G.Pancheri,
N.Paver. 1995. Vol.2, p.451.

\bibitem{lattice}
M.~Crisafulli, V.~Lubicz,
The Second DA$\Phi$NE Physics Handbook. INFN-Laboratori
Nazionali di Frascati. Edited by L.Maiani, G.Pancheri,
N.Paver. 1995. Vol.2, p.515.

\bibitem{pdg}
D.E.~Groom {\it et al.,}
Eur. Phys. J. {\bf C15} (2000) 1.

\bibitem{lande}
R.I.~Dzhelyadin {\it et al.,}
Phys. Lett. {\bf 84B} (1979) 143.

\bibitem{nd1}
S.I.~Dolinsky {\it et al.,}
Sov. J. Nucl. Phys. {\bf 48} (1988) 277.

\bibitem{nd2}
V.B.~Golubev {\it et al.,}
Sov. J. Nucl. Phys. {\bf 41} (1985) 756.

\bibitem{jane}
M.R.~Jane {\it et al.},
Phys. Lett. {\bf 59B} (1975) 103, \\
Errata, Phys. Lett. {\bf 73B} (1978) 503.

\bibitem{e1}
M.N.~Achasov {\it et al.,}
Preprint Budker INP 98-65, Novosibirsk, 1998.

\bibitem{e2}
R.R.~Akhmetshin {\it et al.},
Preprint Budker INP 99-11, Novosibirsk, 1999.


\bibitem{pioeegg_cmd2}
R.R.~Akhmetshin {\it et al.}, Preprint Budker INP 99-97, Novosibirsk,
1999, hep-ex/0011026. To be published in Phys. Lett. B.

\bibitem{vepp}
V.V.~Anashin {\it et al.,} Preprint Budker INP 84-114, Novosibirsk, 1984.

\bibitem{cmddec}
G.A.~Aksenov {\it et al.}, Preprint Budker INP 85-118, Novosibirsk, 1985. \\
E.V.~Anashkin {\it et al.}, ICFA Instr. Bulletin {\bf 5} (1988) 18.

\bibitem{cmd2sim}
E.V.~Anashkin {\it et al.}, Preprint Budker INP 99-1, Novosibirsk, 1999.

\bibitem{rho4pi}
R.R.~Akhmetshin {\it et al.}, Phys. Lett. {\bf B475} (2000) 190.

\bibitem{eidelman}
S.I.~Eidelman,  Proceedings of the Workshop on Physics and Detectors for 
DA$\Phi$NE, Frascati, 1991, p.451. 

\bibitem{hashi}
M.~Hashimoto, Phys. Rev. {\bf D54} (1996) 5611.

\bibitem{solodov}
E.P.~Solodov, Proceedings of the VII Int. Conf. on Hadron Spectroscopy,
Upton, NY, 1997, AIP Conf. Proceed. 432, p.778.
  
\bibitem{lechner}
M.~Lechner, Dissertation Swiss Federal Institute of Technology (ETH)
for Doctor of Natural Science degree, Diss. ETH No. 12866 and ETHZ-IPP
Internal Report 98-5, 1998.



\bibitem{prsnd}
V.M.~Aulchenko {\it et al.,},
Preprint Budker INP 2000-60, Novosibirsk, 2000.

\bibitem{gross}
R.A.~Grossman {\it et al.}, Phys. Rev. {\bf 146} (1966) 993.

\bibitem{dimova}
T.V.~Dimova, private communication.

\bibitem{kroll}
N.M.~Kroll and W.~Wada, Phys. Rev. {\bf 98} (1955) 1355.

\end{thebibliography}
\end{document}